\documentclass[english,aps,prl,superscriptaddress,twocolumn]{revtex4-1}
\usepackage{ae,aecompl}
\usepackage[T1]{fontenc}
\usepackage[latin9]{inputenc}
\usepackage{amstext}
\usepackage{graphicx}
\usepackage{babel}
\begin{document}

\title{Probing 5\textit{f} electronic hybridization in Uranium compounds
via x-ray magnetic circular dichroism}

\author{R. D. dos Reis}

\affiliation{Brazilian Synchrotron Light Laboratory (LNLS), Campinas, SP 13083-970,
Brazil}

\affiliation{Instituto de Fisica Gleb Wataghin, Universidade Estadual de Campinas
(UNICAMP), SP, Brazil}

\author{L. S. I. Veiga}

\affiliation{Brazilian Synchrotron Light Laboratory (LNLS), Campinas, SP 13083-970,
Brazil}

\affiliation{Instituto de Fisica Gleb Wataghin, Universidade Estadual de Campinas
(UNICAMP), SP, Brazil}

\author{D. Haskel}

\affiliation{Advanced Photon Source, Argonne National Laboratory, Argonne, IL
60439, U.S.A.}

\author{J. C. Lang}

\affiliation{Advanced Photon Source, Argonne National Laboratory, Argonne, IL
60439, U.S.A.}

\author{Y. Joly}

\affiliation{Univ. Grenoble Alpes, Inst. NEEL, F-38042 Grenoble, France}

\affiliation{CNRS, Inst. NEEL, F-38042 Grenoble, France}

\author{F. G. Gandra}

\affiliation{Instituto de Fisica Gleb Wataghin, Universidade Estadual de Campinas
(UNICAMP), SP, Brazil}

\author{N. M. Souza-Neto}

\email{narcizo.souza@lnls.br}

\affiliation{Brazilian Synchrotron Light Laboratory (LNLS), Campinas, SP 13083-970,
Brazil}

\affiliation{Advanced Photon Source, Argonne National Laboratory, Argonne, IL
60439, U.S.A.}
\begin{abstract}
We study the spin-dependent electronic structure of UTe and UT$_{2}$Si$_{2}$(T=Cu
and Mn) compounds with a combination of x-ray magnetic circular dichroism
measurements and first principle calculations. By exploiting the presence
of sizable quadrupolar and dipolar contributions to the U L$_{2,3}$-edge
x-ray absorption cross section we are able to provide unique information
on the extent of hybridization between 5\emph{f} and 6\emph{d}/3\emph{d}
electronic states, a key parameter regulating the physical properties
of all actinide materials. Since this information is hardly accessible
to other probes, the new methodology opens up new venues for investigating
this important class of materials.
\end{abstract}
\maketitle
The extent of \emph{f}-state electronic hybridization with valence/conduction
band states is key in defining the physical properties of rare earths
and actinide compounds. In actinide elements \citep{Benedict-HPCRE17,Moore-RMP2009,Santini-AP1999}
the large energy bandwidth of 5\emph{f}-states puts them in an intermediate
scenario between the localized 4\emph{f} states of rare-earths and
the delocalized 3\emph{d} states of transition metals. For example,
the extent of 5\emph{f}-6\emph{d} hybridization is key to address
some still open questions on the itinerant electron behavior in UTe
\citep{Durakiewicz-PRL2004} and UGe$_{2}$ \citep{Saxena-Nature2000}
and the recent interpretation of hastatic order in the heavy-fermion
compound URu$_{2}$Si$_{2}$ \citep{Chandra-Nature2013}. Although
we focus here on magnetic properties\citep{Santini-AP1999}, the \emph{f}-\emph{d}
hybridization also affects intriguing phenomena such as oxidation
states \citep{Glatzel-PRL2013,Souza-Neto-PRL2012}, electronic structure\citep{Caciuffo-PRB2010,Durakiewicz-PRB2004},
pressure induced changes\citep{Heathman-PRB2010,Schilling-JMMM96,Souza-Neto-PRL2009,Souza-Neto-PRL2012,Kolomiets-JKPS2013},
and the character of bonding\citep{Moore-RMP2009}, among others.
A few resonant and non-resonant high resolution x-ray spectroscopy
techniques \citep{Boariu-PRL2013,Caciuffo-PRB2010,Durakiewicz-PRB2004,Durakiewicz-PRL2004,Glatzel-PRL2013,Heathman-PRB2010}
have recently been proposed to study the electronic structure of actinide
compounds showing decisive results in some cases. However, a method
to directly and selectively probe the electronic 5\emph{f}, 6\emph{d}
states and their hybridization has not been available yet, which,
in turn, is crucial for a comprehensive understanding of the unconventional
mechanisms that regulate the physics of 5\emph{f} electrons in actinides.
In addition to that, with the intrinsic dificulties handling actinide
elements due to their toxic and radioactive nature, theoretical work
on actinide compounds has been much more extensive than experimental
studies \citep{Pi-PRL2014,Chandra-Nature2013,Moore-RMP2009,Pickett-PRL2001}.
Therefore the need for a technique capable of directly probing the
relevant electronic states, as well as testing theoretical predictions,
is abundantly clear. 

It is well known that 5\emph{f} states of actinides can be directly
probed by x-ray absorption spectroscopy (XAS) at M$_{4,5}$ (3\emph{d}$\rightarrow$5\emph{f})
or N$_{4,5}$ (4\emph{d}$\rightarrow$5\emph{f}) transitions allowing
determination of orbital and spin magnetic moment via x-ray magnetic
circular dichroism (XMCD) sum rules\citep{Okane-JPSJ2008,Kernavanois-JPCM2001}.
But information on the 6\emph{d} states is absent in these measurements.
By using L$_{2,3}$-edges XAS spectra instead, which involves 2\emph{p}$\rightarrow$5\emph{d},6\emph{d}
and 2p$\rightarrow$4\emph{f},5\emph{f }transitions in the dipolar
and quadrupolar channels\citep{Souza-Neto-PRL2009}, respectively,
information on both 5\emph{f} and 6\emph{d} states (and their hybridization)
can in principle be obtained. However, experimental difficulties such
as the inefficiency to produce circular polarization and the relatively
poor energy resolution available at the high energy range (17 to 21
keV) required for Uranium edges together with the difficulties in
handling radioactive elements at large user facilities have thwarted
efforts to perform careful L$_{2,3}$ XMCD studies on Uranium or any
other actinide compound. Here we report on x-ray magnetic circular
dichroism in the high energy Uranium L$_{2,3}$-edges for UTe and
UT$_{\text{2}}$Si$_{2}$ (T=Cu,Mn) compounds. We demonstrate the
use of this technique in combination with first principle calculations
to directly address the question of electronic hybridization between
5\emph{f} and 6\emph{d} states. We find that an analysis of the relative
amplitudes of (sizable) quadrupolar and dipolar contributions to the
XMCD signal at U L$_{2,3}$-edges allows us to provide previously
unavailable information on this key aspect of the electronic structure
of actinide materials. In addition, we use the element selectivity
of XMCD to show that Uranium sub-lattice carries a induced magnetic
moment above its ordering temperature. The ability to directly probe
experimentally the 5\emph{f}-6\emph{d} hybridization and test validity
of theoretical modes should guide efforts to understand the magnetic
and electronic structures that govern a vast number of actinide compounds
with yet bewildering physical properties.

\begin{figure}
\includegraphics[scale=0.45]{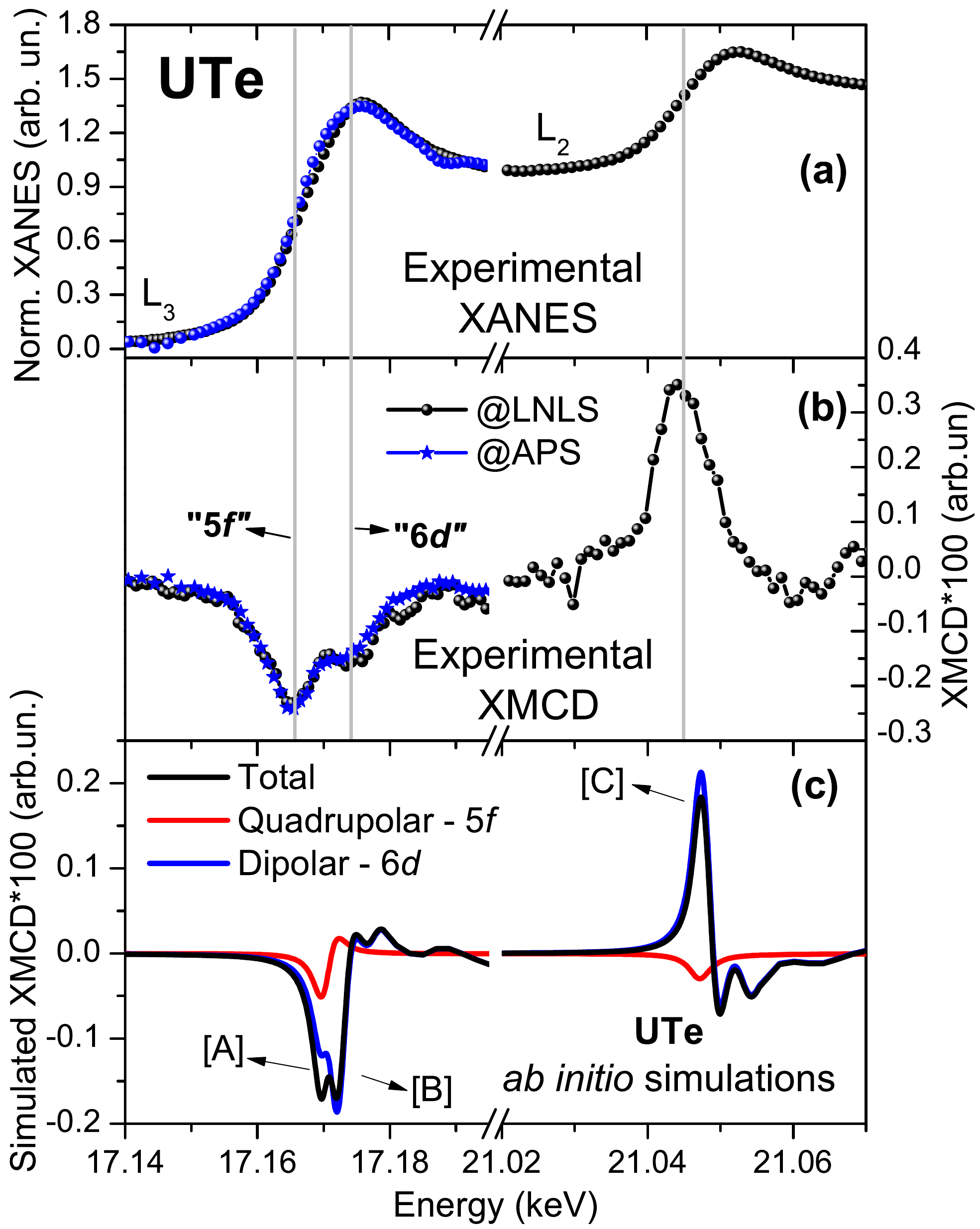}\caption{\label{fig:xmcd-L23-exp-theo}(color online) Uranium L$_{2,3}$-edges
XANES (a) and XMCD (b) spectra measured for the UTe compound at temperature
of 10 K (well below the 104 K ferromagnetic ordering temperature)
and applied field of 0.45 T. In (c) are shown the corresponding \emph{ab
initio} simulations of UTe XMCD spectra using the FDMNES code \citep{Joly-PRB01}.
The features in the experimental spectra are interpreted in view of
dipolar (6\emph{d}) and quadrupolar (5\emph{f}) contributions as discussed
in the text. Lines are guides to the eye. These experiments were performed
at two synchrotron facilities (Advanced Photon Source in Chicago/U.S.A.
and Brazilian Synchrotron Light Laboratory in Campinas/Brazil) with
completely independent samples grown in Germany and Brazil, respectively,
as described in the detailed materials and methods presented in supplemental
materials\citep{supplemental}. This support our findings with equivalent
results from independent experiments and samples.}
\end{figure}

We have selected UTe as the reference compound to benchmark the new
methodology in part because questions remain on the degree of 5\emph{f}
states involvement in its electronic properties. \footnote{Theoretical
studies have proposed different scenarios for UTe based on mixed-valent
regime, the role of orbital hybridization and a collective Kondo state,
and experimental studies have argued in favor of itinerant magnetism
in the 5\emph{f}-electrons \citep{Durakiewicz-PRL2004}}, Uranium
L$_{2,3}$-edge (17.166 keV and 20.95 keV) XMCD measurements on UTe
are presented in Figure \ref{fig:xmcd-L23-exp-theo}, where at the
L$_{3}$ edge two negative peaks (A and B) are evident . With the
support of \emph{ab initio} simulations using the FDMNES code \citep{Joly-PRB01}
shown in Figure \ref{fig:xmcd-L23-exp-theo}c, in which we consider
a 3.0 eV core hole broadening, we argue that one of the two contributions
to the lowest energy XMCD peak (labeled A in the plot) comes from
the 5\emph{f} orbitals through a pure quadrupolar term in the absorption
cross section. In addition to that, a second contribution to the amplitude
of peak A arises from the sizable 5\emph{f}-6\emph{d }hybridizatuib
probed through the dipolar channel, as demonstrated in the theoretical
spectrum of figure \ref{fig:xmcd-L23-exp-theo}c. More details about
the procedure to perform these simulations are given in the supplemental
material \citep{supplemental}, as well as the supporting density
of states determined by LDA+U calculations. The large 5\emph{f}-quadrupolar
term at the L$_{3}$ edge relies on the fact that the x-ray photon
wavelength ($\lambda=0.722\,\mathring{A}$ in this case) is comparable
to the size of the electron orbit \citep{Joly-JPCM2008}. In other
words, the higher energy of the Uranium L-edges (17 keV) enhances
the quadrupolar contribution from the transition Hamiltonian operator
when compared to lower energy L-edges of rare earths\citep{Souza-Neto-PRL2009,Lang-PRL95}
(5 to 10 keV). The dipolar term which essentially probes the 6\emph{d}
orbitals (mainly in peak B) is also present in the same L$_{3}$-edge
spectrum making this an unique tool to directly probe both 5\emph{f}
and 6\emph{d} orbitals in the same experiment. On the other hand,
similarly to the rare earths, the quadrupolar contribution is very
weakly present at the L$_{2}$-edge XMCD due to the combination of
how the spread energy levels and Hamiltonian matrix elements contribute
to the multiple scattering amplitudes in this case \citep{Joly-JPCM2008}
and a very strong spin-orbit coupling influence of Uranium orbitals.
This is verified by the experimental and calculated XMCD spectrum
at the U L$_{2}$-edge (figure \ref{fig:xmcd-L23-exp-theo}) where
the peak C is predominantly a contribution from the dipolar term with
the weak 5\emph{f} quadrupolar term contributing negatively to the
total amplitude. Nevertheless, we could still expect to see some indications
of 5\emph{f}-6\emph{d} electronic hybridization at the L$_{2}$-edge
through the dipolar channel similarly to the L$_{3}$-edge. 

Our findings of the large 5\emph{f} and mixed 5\emph{f}/6\emph{d}
contribution at the L$_{3}$-edge appears to be in contradiction with
the only other report of magnetic study on an actinide L-edge by \citet{Wermeille-PRB98}.
They report on x-ray resonant magnetic measurements of UPd$_{2}$Si$_{2}$
where two peaks appear at the L$_{3}$-edge, both assigned to dipolar
transitions. Their conclusion is based on polarization analysis and
the Q-dependence of the scattered intensities. Although we cannot
directly compare results on different compounds using different techniques,
in what follows we present results on UT$_{2}$Si$_{2}$ compounds
(T=Cu,Mn) with the same U-122 structure as UPd$_{2}$Si$_{2}$ where
experiment and theory show clear evidence for quadrupolar contributions
in the XMCD spectra. 

\begin{figure}
\includegraphics[scale=0.4]{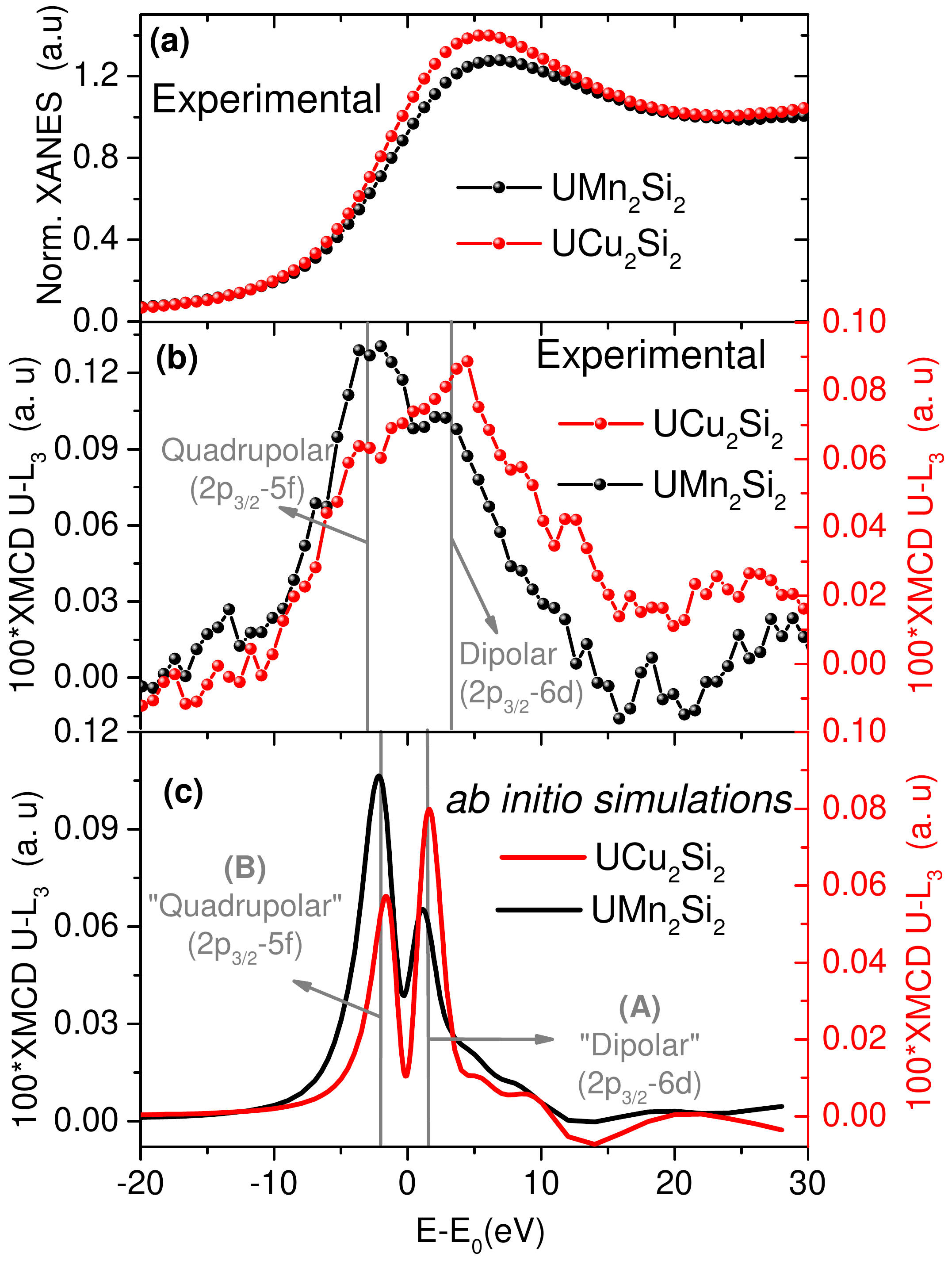}\caption{\label{fig:xmcd_U_UT2Si2} (Color online) Experimental Uranium L$_{3}$
edge XANES (a) and XMCD (b) measurements for UCu$_{2}$Si$_{2}$ and
UMn$_{2}$Si$_{2}$ performed at temperatures of 10 K and 300 K, respectively.
In (c) are shown the corresponding \emph{ab initio} simulations for
both compounds.The difference observed on XMCD experimental data for
the ratio between the intensities of quadrupolar and dipolar intensities
is very well reproduced on the simulated spectra and the explanation
for this behavior is discussed on the main text.}
\end{figure}

\begin{figure}
\includegraphics[scale=0.4]{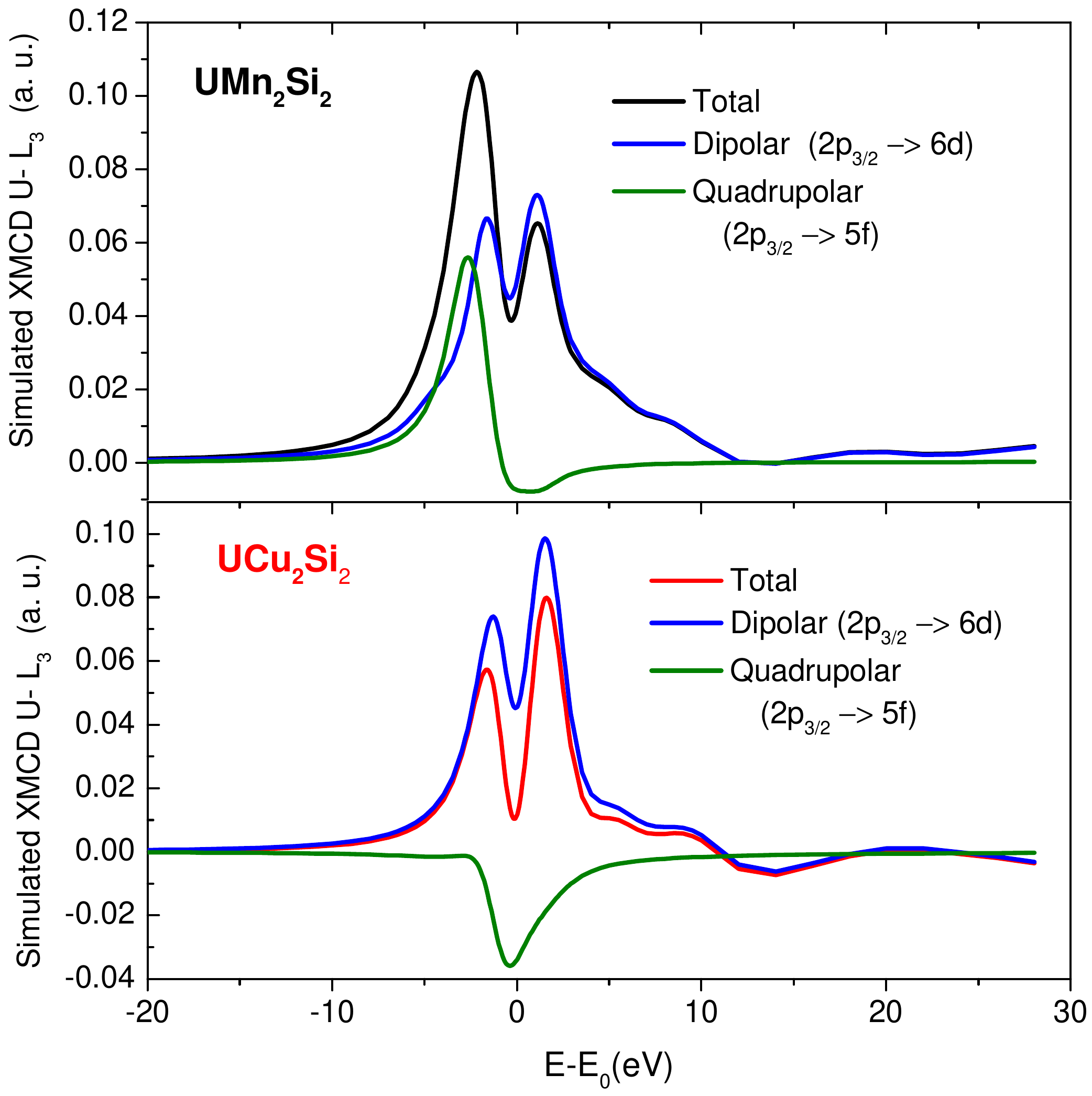} 

\caption{\label{fig:UT2Si2-dos}(Color online) First principle calculations
of each contribution to the XMCD spectra for the UCu$_{2}$Si$_{2}$
and UMn$_{2}$Si$_{2}$ compounds. The quadrupolar contribution has
opposite sign between the two compounds. }
\end{figure}

$\mathrm{UCu_{2}Si_{2}}$ and $\mathrm{UMn_{2}Si_{2}}$ are ferromagnetic
(FM) compounds, with ordering temperatures of $\mathrm{T_{C}}=$103
K and 377 K, of the the family of intermetallics UT$_{2}$X$_{2}$
(1:2:2) with T = transition metal and X = Si or Ge \citep{Mavromaras,SZYTUKA}.
This family presents a wide variety of electronic and magnetic properties,
often times argued to be due to the relationship between magnetism
and the hybridization effects amongst Uranium 5\emph{f }states and
3\textit{d}/4\textit{d}/5\textit{d} states\citep{Mavromaras,Sandratskii19951397}.
The methodology described above is the ideal method to test the validity
of proposed theoretical models for this family of compounds. While
much attention has been given to UCu$_{2}$Si$_{2}$ \citep{Cornelius2008940,Roy1996563,Coqblin_PhysRevB.85.224434},
only a few studies on UMn$_{2}$Si$_{2}$ are reported so far \citep{Chaughule1992385,SZYTUKA}.
The UCu$_{2}$Si$_{2}$ compound is reported to show dual itinerant/localized
character of the 5\textit{f} electrons being a strongly anisotropic
ferromagnet and exhibiting the Kondo effect\citep{Cornelius2008940,Roy1996563}.
In this picture the FM order comes from the part of the U 5\textit{f}
highly localized electrons giving rise to the high T$_{C}$ value\citep{Coqblin_PhysRevB.85.224434}.
On the other hand, UMn$_{2}$Si$_{2}$ orders ferromagnetically \citep{Chaughule1992385,SZYTUKA}
below 377 K for the Mn sublattice and below about 100 K for the U
sublattice. . The electronic structure of this series is characterized
by the relative energy positions of the \textit{d} states of a given
transition metal atom compared with those of the uranium 5\textit{f}
states\citep{Mavromaras}. While in the Mn-compound the 3\textit{d}
and 5\textit{f} states are closer in energy and the \textit{f-d} hybridization
is considered to be the largest among the family of 1:2:2 silicates,
in the Cu compound the 3\textit{d} and 5\textit{f} states are most
distant and therefore the hybridization is almost negligible\citep{Mavromaras,Sandratskii19951397}. 

Figure \ref{fig:xmcd_U_UT2Si2} shows experimental and theoretical
XANES/XMCD spectra of U-L$_{3}$ edges for UCu$_{2}$Si$_{2}$ and
UMn$_{2}$Si$_{2}$. A small difference in the integrated area under
the main XANES peak (so called white-line) is observed between the
two compounds. This indicates that UCu$_{2}$Si$_{2}$ presents a
larger number of unnoccupied states above the Fermi level than UMn$_{2}$Si$_{2}$,
which may be an indication of the former being much less hybridized
than the latter, in agreement with the theoretical predictions\citep{Mavromaras,Morkowski20116994}.
Similarly to the case of UTe, the XMCD data for these two compounds,
shown in Figure \ref{fig:xmcd_U_UT2Si2}(b), present two well defined
peaks. Using the model proposed for UTe, the peak at higher energy
would be related to a dipolar (6\textit{d}) contribution while the
peak at lower energy would be a measure of both quadrupolar contribution
(5\textit{f}) and its overlap with the dipolar contribution due to
5\textit{f}/6\textit{d} hybridization. This scenario is again supported
by first principle calculations of the XMCD spectra as shown on Figure
\ref{fig:xmcd_U_UT2Si2}(c) which reproduce well the experimental
features and differences between the compounds. Interestingly enough,
the amplitude ratio between the two XMCD peaks is opposite for the
two compounds. As supported by the calculations of the dipolar/quadrupolar
contributions shown in Figure \ref{fig:UT2Si2-dos}, this difference
between the two compounds is due to the fact that the dipolar contribution
is almost the same for both compounds while the quadrupolar contribution
has opposite sign between the two materials. Altogether, these results
indicate that while the 5\emph{f} and 6\emph{d} moments are parallel
for $\mathrm{UMn_{2}Si_{2}}$, the same moments are aligned anti-parallel
in the $\mathrm{UCu_{2}Si_{2}}$. This is also consistent with LDA+U
simulations used to estimate the 5\emph{f} / 6\emph{d} moments\citep{supplemental}. 

In addition to this orbital-selective probe of the spin-dependent
empty density of states using XMCD at the U-L$_{2,3}$ edges, it is
also worth commenting on the element selectivity of the technique
to determine that the magnetism of the Uranium for both compounds
seem to have comparable amplitudes. This is surprising considering
the lack of other examples of a magnetic moment on Uranium at ambient
temperature, as we observed here for the UMn$_{2}$Si$_{2}$. Since
a previous neutron scattering study \citep{SZYTUKA} shows ordering
of the Uranium sublattice below 100 K, the sizeable XMCD signal in
the U at 300 K is induced by \emph{f}-\emph{d} hybridization with
the magnetically ordered Mn sublattice. We would expect that at temperatures
below 100 K the low energy peak of the L$_{3}$ edge XMCD would increase
in amplitude even further due to the full magnetic ordering of the
uranium sublattice. This observation of induced quadrupolar contribution
is a quite unusual scenario when compared to rare earths \citep{Lalic-JMMM04}.
In 4\emph{f} systems an induced 5\emph{d} moment would not lead to
an induced 4\emph{f} moment since these orbitals would not strongly
hybridize considering the localized nature of the 4\emph{f} orbitals.
Here, on the other hand, the 5\emph{f}-6\emph{d} hybridize strongly
so an induced 6\emph{d} moment can also induce a 5\emph{f} moment.
This is likely to be taking place as indicated by our theoretical
simulations. 

In summary, we demonstrate that U L$_{2,3}$ XMCD measurements combined
with first principle calculations provide a unique probe of both 5\emph{f}
and 6\emph{d} orbitals, as well as their hybridization, through dipolar
and quadrupolar contributions to the XMCD spectra. Since this hybridization
is key in determining the physical properties of a vast number of
actinide compounds this methodology is bound to improve our understanding
of these correlated electron systems as well as provide an unique
route to test validity of theoretical models. It is noteworthy that
the methodology developed here, using penetrating high-energy x-rays,
can also be applied to study the evolution of \emph{f}-\emph{d} hybridization
under applied pressure providing an unique tool for tuning electron
correlations for discovery of novel phenomena. Finally, we showed
presence of magnetic contributions in Uranium sub-lattice in UMn$_{2}$Si$_{2}$
and UCu$_{2}$Si$_{2}$ compounds, with a clear magnetic signal from
uranium atoms at room temperature in UMn$_{2}$Si$_{2}$, a result
of induced 6\emph{d} and 5\emph{f} moments due to hybridization with
the magnetically ordered Mn 3\emph{d} orbitals. 
\begin{acknowledgments}
We are thankful to Roberto Caciuffo and Gerry Lander for discussions
and comments on the manuscript. We also thank Roberto Caciuffo for
providing one of the UTe samples. Work at Argonne is supported by
the U.S. Department of Energy, Office of Science, Office of Basic
Energy Sciences, under Contract No. DE-AC-02-06CH11357. Work at LNLS
is supported by the Brazilian ministry of science and technology.
RDR and LSIV thank the funding for their Ph.D. fellowships from CAPES
and FAPESP brazilian agencies.
\end{acknowledgments}

\bibliographystyle{apsrev4-1}
\bibliography{Uranium_paper}

\end{document}